# Development of a general equation of state for real molecules in arbitrary regimes of temperature and pressure: I. The hard-core reference system.

**J. F. Kenney**[†]
**Russian Academy of Sciences, Joint Institute of the Physics of the Earth;**
**Gas Resources Corporation,** 11811 N. Freeway, fl. 5,Houston, TX 77032, U.S.A.; JFK@alum.MIT.edu

**Richard J. Petti**
**The MathWorks, Inc.**, 3 Apple Hill Drive, Natick, MA 01760, U.S.A.

Abstract:
   A general equation of state for the hard-body reference system of real fluid has been developed from first principles, statistical mechanical arguments using metric differential geometry to describe the "available volume," $V_0$, and its determining surface, $S_0$, of a hard-body fluid. The rigorous, exact results of scaled particle theory of Reiss *et al.*, which themselves obtain from statistical geometry, have been applied following the extension of Boublík *et al.* for hard bodies of non-spherical shape. The geometric description of the hard-body system can be used with the Boublík equation of state at low and modest densities. At high densities, this geometric description specifies the procedure to specify $V_0$ and $S_0$, notwithstanding that both become multiply-connected.

[Keywords: Statistical mechanics, thermodynamics, hard-body systems, differential geometry, chirality.]

---

[†] author for communication.

# 1. Introduction.

There is a need among physicists working in chemical physics, geophysicists, chemists, and chemical engineers for a rigorous, general equation of state that is valid for molecules of arbitrary shape and in arbitrary regimes of temperature and pressure. The equations of state now commonly used, such as the Redlich-Kwong or Suave equations, have been fitted to convenient experimental measurements, and are therefore invalid in regimes of temperature and pressure outside their range of fitting, or for compounds other than those to which the fittings were made. Others, such as those based on the formalism of the Hard-Chain Theory [HCT], involve invalid assumptions, like the basis of the commonly-used "Prigogine $c$-factor."

Geophysicists need to describe and predict the behavior of chemically reactive systems in regimes of temperature and pressure found in the depths of the Earth. There exist very few measured data for minerals or compounds in the thermodynamic regimes of the Earth's depths. Therefore, valid geophysical predictions must depend upon the validity of the equation of state used.

The number of synthesized chemical compounds already exceeded 6 million at the end of the last century.[1] With the developments of chemical synthesis, major pharmaceutical companies expect to synthesize approximately 100,000 new compounds yearly. Each time a new compound is synthesized and determined to be of commercial value, the scientists and chemical engineers involved are confronted with the tasks of separation, purification, and production on an industrial scale in quantities of tons of a material previously known only in quantities of grams (or kilograms), and for which the physical properties of the compound, such as its density as a function of pressure and temperature, are almost always unknown. When the compound(s) of interest involves a mixture, its physical properties must be predictable as functions of the composition of the system.

These tasks require a general equation of state which must involve no fitted parameters or additive constants. This equation of state must require only minimal physical properties of the individual molecule which can be measured separately, e.g., such as the molecular geometry and the shape of its electronic cloud which can be obtained from the x-ray scattering form factor. Such equation of state must be derived from first-principles, quantum statistical mechanics arguments in order hold for arbitrary regimes of temperature and pressure.

This article reports the first part of the development of a rigorous equation of state, derived from first-principles, statistical mechanics arguments, and applicable to real molecules of non-convex shape in general regimes of temperature and pressure. This analysis uses the technique of the factored partition function enunciated first by Bogolyubov,[2] later independently by Feynman,[3] and developed substantially by Yukhnovskii.[4-7] The factored canonical partition function applies a "reference system" together with a subsidiary partition function, such that the quantal canonical partition function has the form:

$$Q(V,T,\{n_i\}) = \text{trace}\left\langle Q^{\text{ref}}(V,T,\{n_i\})\, Q^{\text{int}}(V,T,\{n_i\};\boldsymbol{r}^{\text{ref}})\right\rangle = \text{trace}\left\langle Q^{\text{ref}}\right\rangle \text{trace}\left\langle Q^{\text{int}}\right\rangle. \qquad (1)$$

The first factor on the right side of equation (1) is the partition function for the reference system, which subsumes the Ideal Gas partition function together also with that resulting from the short-range, repulsive, hard-core component of the intermolecular potential. The second factor accounts for the long-range, attractive, weak component of the intermolecular potential, often designated the van der Waals component.

The use of the factored partition function in equation (1) represents a generalization of the adiabatic formalism used in molecular and solid state physics whereby the quantal state vector of the system is described in terms of the factors of a state vector for the electronic system which is a parametric function of the nuclear coordinates, and another for the nuclear coordinates:

$$\left| \Upsilon\left(\vec{x}_1, \vec{x}_2, \cdots, \vec{x}_N, \vec{X}_1, \vec{X}_2, \cdots, \vec{X}_N\right) \right\rangle = \left| \Psi\left(\vec{x}_1, \vec{x}_2, \cdots, \vec{x}_N; \{\vec{X}_i\}\right) \right\rangle \left| \Phi\left(\vec{X}_1, \vec{X}_2, \cdots, \vec{X}_N\right) \right\rangle. \quad (2)$$

In equation (2), the lower case variables represent the electronic coordinates, and the upper case ones the nuclear coordinates (typically, the phonon system). The factoring process involved in equations (1) and (2) does not introduce, *per se*, any approximation or imprecision; only such specific approximations as may be introduced to either factor introduces imprecision.

The factored partition function allows the equation of state to be described separately in terms of the different contributions to the intermolecular potential. The system pressure is given by:

$$p = -\left(\frac{\partial F}{\partial V}\right)_{T,\{n\}} = -k_B T\left(\frac{\partial \ln Q(V,T,\{n_i\})}{\partial V}\right)_T = -k_B T\left(\frac{\partial \ln Q^{ref}}{\partial V}\right)_T - k_B T\left(\frac{\partial \ln Q^{int}}{\partial V}\right)_T = p^{ref} + p^{int}, \quad (3)$$

in which $F$ is the Helmholtz free energy, and $k_B$ is Boltzmann's constant; $p$, $V$, and $T$ are, respectively, the system pressure, volume, and absolute temperature.

This article addresses the development of the partition function, and its related equation of state, for the hard-core reference system: $Q^{ref}$, $p^{ref}$. The reference system used here is that of the fluid of hard-body particles. The factor in the partition function which describes the effects of the weak, attractive, long-range intermolecular interaction, and the component of pressure which obtains from it, will be the subject of a following article.

The short-range, repulsive, hard-core component of the intermolecular potential obtains from the Jordan-Pauli exclusion principle which (as its name implies) dictates that every molecule occupies a volume of space into which another molecule cannot be injected. This most fundamentally quantal property determines the motivating perspective of the hard-body fluid and its analysis by statistical geometry.[8-10]

In a previous article,[11] on the high-pressure physics of multi-component systems, the effects of geometry on the evolution of hard-body systems at high densities were examined generally. In another previous article,[12] the effects of geometry upon the evolution of many-body systems of real molecules at high densities were examined for the very restricted case of mixtures of molecules of equal volume whose geometries were identical except for being chiral enantiomers. In a more recent article,[13] the effects of geometry upon the evolution of the real, chemically-reactive, many-body, hydrogen-carbon [H-C] system were analyzed at high densities in order to determine the thermodynamic regimes of pressure at which the compounds heavier than methane can be generated spontaneously. This article analyses the hard-body fluid for particles of general convex shape and of a representative number of non-convex shapes, and relates their molecular geometry to the sub-volume of the system into which another particle can be injected, designated the "available volume."

This paper is organized into nine sections, as follows.

Section 2 reviews the general geometric equation of state for fluids, which was first written by Boltzmann and later derived independently by Speedy[14-16] and by Reiss and Hemmerich.[17]

Section 3 reviews the derivation by Reiss *et al*[8, 18, 19] of the equation of state of a hard-sphere fluid by scaled particle theory from the constraints of statistical geometry.

Section 4 briefly revisits the geometric description of a hard-body particle as developed by Steiner and Kihara[20-22] for convex particles.

Section 5 introduces the development by Boublík[23] *et al.*[24] of the extension of scaled particle theory to hard-body molecules of arbitrary, but still convex, geometries, and the use of the Steiner-Kihara parameters in scaled particle theory.

Section 6 returns to the concept and use of the "surface of exclusion" into which no particle may be inserted, as was introduced by Boltzmann and used by Reiss *et al.*, and introduces the formalism of metric differential geometry to describe the geometry of hard-body particles. The restriction that the hard-body molecules be convex is removed. The surface parameter used by Steiner and Kihara, $S^{S-K}$, is replaced by the pertinent "exclusionary surface" determined by the particular geometry of the molecule(s) considered. The Steiner-Kihara radial parameter, $R^{S-K}$, is generalized to the mean extrinsic curvature of the bounding surface. The use of metric differential geometry of embedded surfaces to describe the geometry of hard-body particles removes the constraint of convexity, while preserving the exactness of the scaled particle theory. Definition of the mean extrinsic curvature locally requires only that the surface of exclusion be oriented and piecewise twice differentiable; computing the integral requires that the surface be bounded; the surface may be otherwise irregular, non-convex, and even multiply-connected.

Section 7 catalogues computed values of the geometric parameters *R*, *S*, and *V* for some convex and non-convex surfaces with saddle-shaped regions. For all of convex shapes, the respective values of the Steiner-Kihara parameters, $R^{S-K}$, $S^{S-K}$, and $V^{S-K}$, are identical those developed by the differential geometry of embedded surfaces. The method of Kihara for computing *R* is extended to non-convex shapes.

Section 8 catalogues computed values of the geometric parameters *R*, *S*, and *V* for several convex and non-convex surfaces that are typical of the shapes of molecules or of excluded volumes of molecules. The Boublík parameter of asphericity, $\boldsymbol{a}^{\text{Boublík}} = RS/3V$, is shown to be equal for a non-convex hard-body molecule, in many cases, to that for a convex molecule possessing certain geometric similarities. Because the Boublík parameter of asphericity describes the geometric properties of molecules in scaled particle theory, this useful property justifies the modeling of (often complicated) non-convex molecules by simpler convex ones in thermodynamic analyses.

Section 9 compares some thermodynamic quantities computed from this theory with empirical measurements. The computed values of the Boublík parameter of asphericity, $\boldsymbol{a}$, and of the hard-body parameter representing the volume of exclusion agree with the experimentally measured geometric properties of the normal alkane molecules, $C_nH_{2n+2}$.

## 2. The general equation of state of a fluid.

The general equation of state which directly determines the pressure, temperature, and density of a fluid is given by statistical geometry in terms of the "available volume" of the fluid, $V_0$, and the boundary of the space determined by that volume, $S_0$.[8, 25-28] The available volume, $V_0$, is defined as the averaged space available for insertion of another particle into the system; and its surface, $S_0$, is the corresponding area which separates that available space from the _un_available volume, $(V - V_0)$.

The pressure, $p$, given by statistical geometry is:

$$p\boldsymbol{b} = \boldsymbol{r}\left(1 + \frac{\tilde{R}}{3}\frac{S_0(\boldsymbol{r})}{V_0(\boldsymbol{r})}\right) \quad (4)$$

in which $\boldsymbol{b} = 1/k_BT$; and $\boldsymbol{r} = N/V$, and $\tilde{R}$ is a parameter with the dimension of length which measures the averaged distance from a point to the surface, $S_0$, that determines the volume into which no part of another particle can intrude.

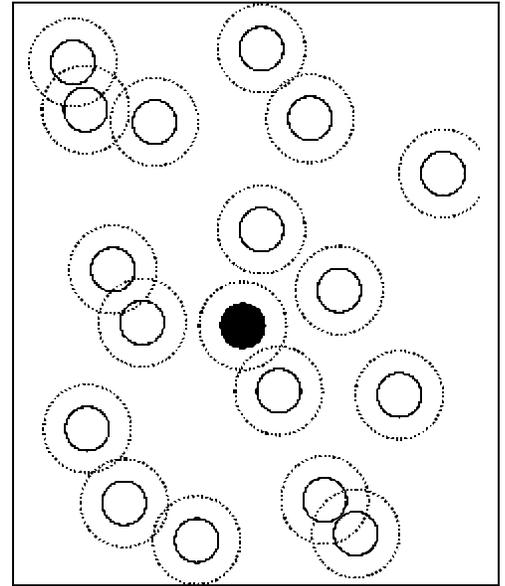

**Fig. 1 Schematic, two-dimensional representation of a "snap-shot" of the dilute, hard-sphere gas. The solid circles represent hard-sphere molecules; the concentric dotted lines outline the region about each particle from which the *center* of another sphere is excluded. A "test particle" is indicated in black.**

This general equation of state, (4), relates the thermodynamic properties of a fluid to an intrinsic geometric property of its component particles, $\tilde{R}$, and to the geometry of an abstract surface, $S_0$, which in turn depends on both the intrinsic geometries of its components and also the system density. For a hard-sphere gas in regimes of dilute to moderate density, $\tilde{R}$ is the radius of the individual hard spheres. For a fluid of convex hard-body particles, $\tilde{R}$ is the radial parameter as defined by Steiner and Kihara. As is shown in sections below, generally, and for particles that are not necessarily convex nor even simply-connected, the parameter $\tilde{R}$ is the "extrinsic curvature," the integral over the surface $S_0$ of the extrinsic curvature tensor of Riemann, $EK$, (called, sometimes, the "second fundamental form").

The equation of state, (4), has an interesting history. In the late 1970's, equation (4) was derived by Speedy[14-16] for a hard-sphere gas, applying arguments that used a lattice-gas formalism which were subsequently passed to a continuum. Several years later, Reiss[17] derived equation (1) directly and rigorously using scaled particle theory. In 1984, Montroll[29] pointed out that this equation of state was known to Boltzmann.

Equation (4) obtains from general arguments of statistical geometry, and is not dependent upon any specific approximate formalism, such as the Kirkwood or Percus-Yevick equations. Equations (5) and (4) are not restricted to hard-body fluids, and hold also when there exists an attractive component to the intermolecular potential.

The chemical potential, $\boldsymbol{m}$, given by statistical geometry is:

$$\boldsymbol{mb} = \ln\left(\frac{N\boldsymbol{l}^D}{V_0}\right) \quad (5)$$

where $\boldsymbol{l}^D = h/\sqrt{(2\pi mk_BT)}$ is the thermal de Broglie wavelength and $m$ the molecular mass. This equation, for a real fluid, has identically the form as that for the chemical potential of an ideal gas, but with the available volume, $V_0$, replacing the system volume, $V$.

Statistical geometry establishes that the configurational factor of the canonical partition function for a system of $N$ hard-body particles is:

$$Z_N = \prod_{n=0}^{N} V_0(n) \tag{6}$$

where $V_0(n)$ is the available free volume into which the center of another particle can be injected for a system of for a system of $n$ particles. The surface which bounds this available free volume is designated $S_0(n)$ and separates that volume from the "excluded volume," $(V - V_0(n))$. Two consequences of the partition function for the hard-body system are the explicit expressions for the system pressure and the component chemical potentials.

Equations (5), (4), and (6) are derived explicitly in Appendix A for a hard-sphere gas, following Reiss.[17, 30] In order to render the exposition clear, a schematic "snap-shot" of a hard-sphere gas is given in Fig. 1. One should note in Fig. 1 that, although the solid lines representing the surfaces of hard spheres can never overlap, the concentric dotted lines representing the surfaces from which the centers of another sphere are excluded can do so; and one may note the momentary two-body collision of the pair of particles in the lower right.

### 3. The development by Scaled Particle Theory [SPT] of the equation of state specified by statistical geometry.

The first step of scaled particle theory determines the conditional probability, $P(r)$, that no particle might exist within a spherical region of radius $r$. Consider a fluid of $N$ particles in volume $V$, characterized by the density $\rho = N/V$. With the definition of $P(r)$, the probability that no particle exists in the expanded region of radius $r + dr$ is:

$$P(r+dr) = P(r) + \frac{dP(r)}{dr} dr = P(r)\left(1 - 4\pi r^2 \rho G(r) dr\right). \tag{7}$$

In equation (7), the correlation function $G(r)$ is defined such that $\rho G(r)$ measures the particle density at $r$. Therefore the conditional probability that there is a particle within the spherical shell $dr$ is $4\pi r^2 \rho G(r)$. Therefore, the conditional probability, $P(r)$, must satisfy the differential equation,

$$\frac{1}{P(r)} \frac{dP(r)}{dr} = -4\pi r^2 \rho G(r), \tag{8}$$

such that

$$P(r) = \exp\left(4\pi\rho \int_0^r \lambda^2 G(\lambda) d\lambda\right). \tag{9}$$

In equation (9), the limit condition $P(0) = 1$ has been used. In equations (8) and (9), the particles have been assumed to be spherical, and the angular part of the integration has been subsumed by the factor $4\pi$; for aspherical entities, such factors assume the qualities of averaged values.

To be noted particularly is that equation (9) involves no physical constants, neither Boltzmann's, nor Planck's constant, nor the speed of light, nor any other; nor does it contain temperature explicitly. The reason for this attribute is that equation (9) is an expression of statistical geometry, and is independent both of thermal variables and any physical constant which might set a unique scale to its regime of validity. Thus, the conditional probability, $P(r)$ given by (9), describes the distribution of macroscopic grains and voids in such as po-rous and fractured media, no less than molecules in a gas, - and arguably, with equal validity, nuclear particles or galaxies. Scaled particle theory has indeed been applied to describe the movement of macroscopic fluids in porous and fractured media.[27]

The name scaled particle theory has been given to this exercise is statistical geometry. In Appendix B, the derivation of the equation of state for a hard-sphere gas, (10), is given explicitly:

$$p\boldsymbol{b} = \boldsymbol{r}\left(\frac{1+\boldsymbol{h}+\boldsymbol{h}^2}{(1-\boldsymbol{h})^3}\right), \quad \text{in which } \boldsymbol{h} = \frac{\boldsymbol{p}}{6}\boldsymbol{r}\boldsymbol{s}^3. \tag{10}$$

This scaled particle theory equation of state for a hard-sphere fluid, (10) represents one of the (very) few exactly solvable models in statistical mechanics. Equation (10) is often referred to, inappropriately, as the Percus-Yevick equation. Correctly, equation (10) should be called the Reiss-Frisch-Lebowitz equation, for it was derived and published,[8] a few years before the Percus-Yevick solution, and owes nothing to it.

Carnahan and Starling[31] later developed an equation of state for the hard-sphere gas by summing a series fitted to the values of the first seven virial coefficients obtained from Monte Carlo simulations which is identical to equation (10) with an additional cubic term in the packing fraction, $\boldsymbol{h}$:

$$p\boldsymbol{b} = \boldsymbol{r}\left(\frac{1+\boldsymbol{h}+\boldsymbol{h}^2-\boldsymbol{h}^3}{(1-\boldsymbol{h})^3}\right) = \boldsymbol{r}\left(1+\frac{2\boldsymbol{h}(2-\boldsymbol{h})}{(1-\boldsymbol{h})^3}\right). \tag{11}$$

Because of its wide use, the SPT-Carnahan/Starling equation, (11), will be employed in this paper.

# 4. The geometric description by Steiner and Kihara of a system of convex, aspherical, hard-body particles.

The first systematic and rigorous analysis of the thermodynamics of aspherical hard-body systems in terms of their geometry were carried out by Kihara and Steiner.[20, 21] Kihara and Steiner described the geometry of hard-body particles in terms of three parameter functionals, $R$, $S$, and $V$, which they determined from the support function that describes the volume and surface generated by the rolling of one hard body around the surface of another, in all possible orientations. Steiner and Kihara used a procedure involving the parallel planes, separated by the vector of magnitude $r$ normal to the surface of the hard body, as indicated schematically in Fig. 2. Steiner and Kihara allowed the second convex hard body to roll, in all possible orientations, against a point of contact on that parallel plane, and thenafter allowed that plane to touch the first convex hard body at all points on its surface. Finally, the magnitude of the vector, $u$, normal to the surface of the first convex hard body to the parallel plane(s) to go to zero.

The parameter functional $R$ represents the averaged half-distance of the centers of two particles, and $S$ the surface area determined by $R$; in this instance, $V$ does *not* necessarily represent the partial volume of the hard-body particle itself but the effective volume determined by the support function. In order to assist visualization of the support function and the mathematical entities which enter the Steiner-Kihara equations, a two-dimensional representation of a typical element of the development of $R$, $S$, and $V$, is shown in Fig. 2. The individual functionals $R$, $S$, and $V$, are defined by the equations:

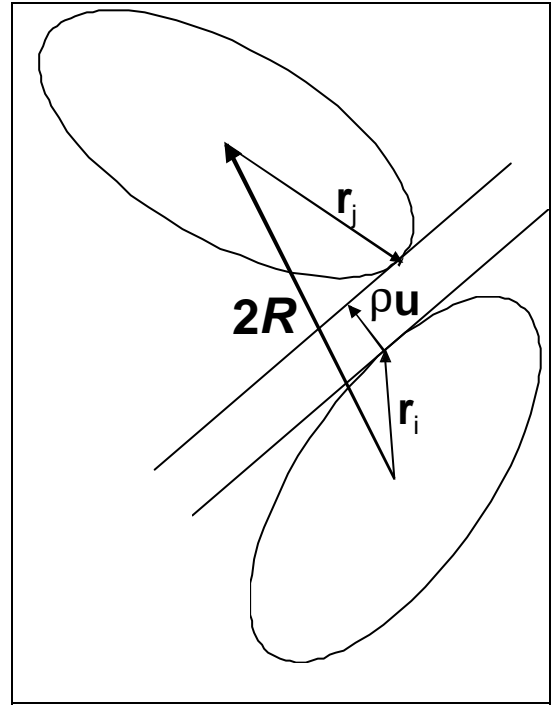

**Fig. 2 The geometric parameters used by Kihara to determine $R$, $S$, and $V$ in equations (12).**

$$\left. \begin{aligned} R &= \frac{1}{4\pi} \int_0^\pi \int_0^{2\pi} \vec{r} \cdot \left( \frac{\partial \vec{u}}{\partial \theta} \times \frac{\partial \vec{u}}{\partial \phi} \right) d\theta d\phi \\ S &= \int_0^\pi \int_0^{2\pi} \vec{u} \cdot \left( \frac{\partial \vec{r}}{\partial \theta} \times \frac{\partial \vec{r}}{\partial \phi} \right) d\theta d\phi \\ V &= \frac{1}{3} \int_0^\pi \int_0^{2\pi} \vec{r} \cdot \left( \frac{\partial \vec{r}}{\partial \theta} \times \frac{\partial \vec{r}}{\partial \phi} \right) d\theta d\phi \end{aligned} \right\} \quad (12)$$

In equations (12), $\vec{u}(\theta, \phi)$ is the unit vector in the direction of the normal of the supporting plane, and $\vec{r}(\theta, \phi)$ is the vector from the origin to the contact point of the convex body with the supporting plane, as indicated in Fig. 2; the angles $\theta$ and $\phi$ are polar angles. As will be shown below, the Steiner-Kihara parameter for the radius, $R$, is the mean extrinsic curvature of Riemann.

## 5. The Boublík equation of state for systems of convex hard-body particles.

Pavlícek, Nezbeda, and Boublík[24] analyzed the equations developed by scaled particle theory for aspherical, convex, hard bodies. Pavlícek *et al.* recognized that the equations for the two-body correlation function, and its derivative, as appear in equations (7) and (8), must depend upon the relative orientations of the two hard-bodies, such that $G = G(r,q,f)$. They integrated out the angular variables in the resulting equations following the technique of Steiner and Kihara, and developed a rigorous equation of state for aspherical (but convex) hard-body particles which involves only the averaged geometric parameters $R$, $S$, and $V$. Boublík (1981)[18, 32] further refined the extension of scaled particle theory for mixtures of hard-body particles of different sizes and shapes and developed the following analytical expression for the pressure of the hard-body fluid in terms of their respective geometric parameters $R$, $S$, and $V$:

$$b p = r \left[ 1 + \left( \frac{h}{(1-h)} + \frac{rs}{r(1-h)^2} + \frac{qs^2(1-2h) + 5rsh^2}{3r^2(1-h)^3} \right) \right] = b \left( p^{IG} + p^{hc} \right)^{Boublik}. \tag{13}$$

In equation (13), the geometric variables of composition, $r$, $q$, $s$, and $u$ are:

$$\left. \begin{array}{lll} r = r \sum_i x_i R_i & q = r \sum_i x_i R_i^2 & s = r \sum_i x_i S_i \\ u = r \sum_i x_i V_i = h & & \end{array} \right\}, \tag{14}$$

in which the sets $[R_i, S_i, V_i]$ are, respectively, the geometric parameters of the support function of the *i*-th component, $R_i$, $S_i$, $V_i$, as set forth in the previous section. In the following sections, the geometric parameter $S$ will be shown to be the surface of the particle (or cluster of particles) from which no part of another particle can intrude, and $R$ will be shown to be the mean exterior curvature of Riemann generated by that surface; and the geometric parameter $V$ is the volume subtended by $S$. By such identification, the restriction of the Boublík equation to convex hard-bodies will be removed, such that scaled particle theory can be used confidently with hard-body particles of general shape.

The geometric variables $r$, $q$, and $s$ in equations (14) may be rewritten so as to separate their respective dependences upon density and geometry by introducing the new variables $\tilde{R}$, $\tilde{Q}$, $\tilde{S}$ and $\tilde{V}$ such that,

$$\left. \begin{array}{lll} r = r\tilde{R} & q = r\tilde{Q} & s = r\tilde{S} \\ u = r\tilde{V} = h & & \end{array} \right\}. \tag{15}$$

The new variables $\tilde{V}$, $\tilde{S}$ and $\tilde{V}$ may be considered as averaged values for the system's geometric parameters, statistically weighted according to composition; similarly, $\tilde{Q}$ may be considered as the compositionally weighted system average of $\tilde{R}^2$. When $\tilde{R}$, $\tilde{Q}$, $\tilde{S}$ and $\tilde{V}$ are used, the Boublík equation can be written as:

$$\frac{pb}{r} = \frac{1}{1-h} + \frac{(\tilde{R}\tilde{S}/\tilde{V})h}{(1-h)^2} + \frac{\tilde{Q}\tilde{S}^2(1-2h) + 5\tilde{R}\tilde{S}h^2}{3(1-h)^3} \tag{16}$$

The Boublík equation, (13), can be further simplified by introducing the following geometric functions, $a$ and $g$:

$$\boldsymbol{a} \equiv \frac{\left(\sum_i x_i \tilde{R}_i\right)\left(\sum_i x_i \tilde{S}_i\right)}{3\left(\sum_i x_i \tilde{V}_i\right)}; \qquad \boldsymbol{g} \equiv \frac{\left(\sum_i x_i \tilde{R}_i^2\right)}{\left(\sum_i x_i \tilde{R}_i\right)^2} \quad \right\}. \tag{17}$$

The geometric parameter $\boldsymbol{a}$ in equations (17) is the multi-component analogue of the Boublík parameter of asphericity for a single-component fluid, $\boldsymbol{a}^{\mathrm{B}} = RS/(3V)$, and may be interpreted as the weighted degree of asphericity for the system. The parameter $\boldsymbol{g}$ in equations (17) has no analogue in a single-component fluid, for which it is always equal to unity; $\boldsymbol{g}$ might be interpreted as a parameter of interference which measures the degree of difference in the mean component dimensions.

When the definitions (17) are used, the Boublík equation, (13), can be written in a simple form as,

$$\boldsymbol{b} p = \boldsymbol{r}\left[1 + \frac{c_1 \boldsymbol{h} + c_2 \boldsymbol{h}^2 + c_3 \boldsymbol{h}^3}{(1-\boldsymbol{h})^3}\right], \tag{18}$$

in which

$$c_1 = 3\boldsymbol{a} + 1, \qquad c_2 = 3\boldsymbol{a}(\boldsymbol{ag} - 1) - 2, \qquad c_3 = 1 - \boldsymbol{a}(6\boldsymbol{ag} - 5). \tag{19}$$

The parameters, $c_1$, $c_2$, and $c_3$ in equations (19) depend only upon the combined geometries of the respective support functions of the molecular components, as expressed by the geometric shape parameters, $\boldsymbol{a}$ and $\boldsymbol{g}$.

Similarly, when the identities (19) are used, the equations for the contribution of the reference system to the Gibbs free enthalpy becomes also a simplified function of the packing-fraction, $\boldsymbol{h}$, and the two geometric variables of the mixture, $\boldsymbol{a}$ and $\boldsymbol{g}$.

$$\boldsymbol{b} G = \boldsymbol{b}\sum_i n_i \boldsymbol{m}_i = N\left[\sum_i x_i \ln\left(\frac{(n_i \boldsymbol{l}_i^3)}{V}\right) + \boldsymbol{h}\left(\frac{I + J\boldsymbol{h} + K\boldsymbol{h}^2}{(1-\boldsymbol{h})^3}\right) - c_3 \ln(1-\boldsymbol{h})\right], \tag{20}$$

in which

$$\begin{aligned} I &= 2c_1 - c_3 &= \boldsymbol{a}(6\boldsymbol{ag} + 1) + 1 \\ J &= -\frac{1}{2}(3c_1 - 3c_2 - 5c_3) &= -\frac{1}{2}(\boldsymbol{a}(21\boldsymbol{ag} - 7) + 4) \\ K &= \frac{1}{6}(3c_1 - 3c_2 - 3c_3) &= \frac{1}{2}\boldsymbol{a}(3\boldsymbol{ag} + 1) + 1 \end{aligned} \right\}. \tag{21}$$

Comparison of equations developed from scaled particle theory, (18) and (20), respectively, with the general equation of state for a fluid, (4), and of its Gibbs free enthalpy, (5), admits a direct connection between the scaled particle theory equation of state and the inverse ratio of the available volume and its surface, $S_0/V_0$, to the packing-fraction, $\boldsymbol{h} = N\boldsymbol{u}/V$:

$$\frac{\tilde{R}}{3}\frac{S_0(\boldsymbol{s})}{V_0(\boldsymbol{s})} = \frac{c_1 \boldsymbol{h} + c_2 \boldsymbol{h}^2 + c_3 \boldsymbol{h}^3}{(1-\boldsymbol{h})^3}. \tag{22}$$

Thus equation (22) provides a direct relationship between the geometric properties of the component molecules and the geometric properties of the surface of exclusion, $S_0$, and its subtended volume, $V_0$, and the com-

ponent parameter, R. In the following sections, the general geometric descriptors of hard-body particles, R, $S_0$, and $V_0$ are developed using metric differential geometry.

## 6. The general description of hard-body particles and the development of the geometric parameters R, S, and V using metric differential geometry.

In this section, the description of the geometry of a hard-body particle is developed in terms of the parameters R, S, and V using metric differential geometry. The resulting parameters R, S, and V so developed are independent of the choice of coordinate systems, both in 3-space and also on the surface of the body, and describe generally hard bodies that are convex, non-convex, and even multiply-connected.

Consider a body **B** that is a smooth 3-manifold-with-boundary embedded in Euclidean 3-space. Its boundary $\partial B$ is a 2-surface embedded in 3-space. First, universal coordinate-invariant expressions for the area of $\partial B$ and the volume of **B** are reviewed; thenafter the parameter R is developed.

### 6.1. Coordinate-invariant expressions for the surface, S, and the volume, V.

Let $x^i$ ($i = 1,2,3$) be coordinates on Euclidean 3-space. Examples of allowable coordinate systems include: (1), Cartesian coordinates, $x^i = (x,y,z)$; (2), spherical coordinates, $x^i = (r,\theta,\phi)$. Let $g_{ij}$ = metric tensor on 3-space ($i,j = 1,2,3$). The metric tensor, $g_{ij}$, captures the information about lengths and inner products of vectors, where the inner product of two vectors **u** and **v** is:

$$\mathbf{u} \cdot \mathbf{v} = u^i g_{ij} v^j = \sum_{(i,j)=1}^{3} u^i g_{ij} v^j . \tag{23}$$

(Throughout this paper the Einstein index notation is used whereby the same letter, when used as an upper and lower index in the same tensor or product of tensors, indicates that the letter index is summed over its entire range.) The metric tensor is symmetric, $g_{ij} = g_{ji}$, because inner products are symmetric, $\mathbf{u} \cdot \mathbf{v} = \mathbf{v} \cdot \mathbf{u}$.
Here follow familiar examples of metric tensors:

$$g_{ij}^{\text{Cartesian}} = d_{ij} = \begin{bmatrix} 1 & 0 & 0 \\ 0 & 1 & 0 \\ 0 & 0 & 1 \end{bmatrix} \quad ; \quad g_{ij}^{\text{spherical}} = \begin{bmatrix} 1 & 0 & 0 \\ 0 & r^2 & 0 \\ 0 & 0 & r^2 \sin^2(\theta) \end{bmatrix}. \tag{24}$$

Theorem 1: The volume form on 3-space is:
$$dVol = \sqrt{\det(g_{ij})} dx^1 dx^2 dx^3 . \tag{25}$$

Examples of volume forms include: (1), in Cartesian coordinates, $dVol = 1 \cdot dxdydz$; (2), in spherical coordinates, $dVol = |r^2 \sin(\theta)| dr d\theta d\phi$.

The volume of B is:
$$V = \iiint_b \sqrt{\det(g_{ij})} dx^1 dx^2 dx^3 . \tag{26}$$

The embedded 2-surface, $\partial B$, is expressed in coordinate-invariant form as follows. Define a portion of $\partial B$ by three functions: $x^i(u,v)$, where $u$ and $v$ form a local coordinate system on $\partial B$. A more compact notation is to denote the coordinates $(u,v)$ on $\partial B$ by $w^a$, ($a = 1,2$), where $w^1 = u$, $w^2 = v$. The metric $g_{ij}$ on 3-space defines lengths and angles on the surface $\partial B$; therefore, it induces a metric tensor on $\partial B$. Let $h_{ab}$ be this metric tensor on $\partial B$.

Theorem 2: The metric tensor induced on $\P B$ is:

$$h_{ab} = x^i_{,a} g_{ij} x^j_{,b} = \sum_{i,j=1}^{3} \frac{\partial x^i}{\partial w^a} g_{ij} \frac{\partial x^j}{\partial w^b}. \tag{27}$$

Proofs of these theorems may be found in standard references on differential geometry, e.g., Struik[33] pp 73-77, Kryszig[34] pp 118-120.

The area form on the surface, $\P B$, and the area of $\P B$ are given, respectively, by:

$$\left. \begin{array}{l} dA = \sqrt{\det(h_{ab})} dw^1 dw^2 \\ S = \int dA = \iint \sqrt{\det(h_{ab})} dw^1 dw^2 \end{array} \right\}. \tag{28}$$

For example, on a sphere of radius, $r$, with angular coordinates $q$ and $f$, the metric tensor is:

$$h_{ab} = \begin{bmatrix} r^2 & 0 \\ 0 & r^2 \sin^2(q) \end{bmatrix}; \tag{29}$$

the area form, $dA$, and area, $S$, of $\P B$ are, respectively: $dA = r^2 \sin(q) \, dq \, df$; $S = \int dA = 4\pi r^2$.

This formulation of the geometry of Euclidean space and embedded surfaces is coordinate-invariant (but not coordinate-free), in the sense that there are straightforward methods for transforming all these constructions to another coordinate system so that lengths, angles, areas, and volumes are the same in the new coordinates. Let $y^a$ ($a = 1,2,3$) be a second set of coordinates on 3-space. The keys to transforming between coordinate systems are: the coordinate functions $y^a = y^a(x^i)$ and their inverse functions $x^i = x^i(y^a)$; and the Jacobian matrix of the partial derivatives $\partial x^i / \partial y^a$ and its inverse matrix $\partial y^a / \partial x^i$.

The metric tensor and the volume forms transform as follows when converting from coordinates $x^i$ to $y^a$.

$$g_{ij} \to g'_{ab} = \sum_{i,j} \frac{\partial x^i}{\partial y^a} g_{ij} \frac{\partial x^j}{\partial y^b}. \tag{30}$$

such that,

$$\det(g_{ij}) \to \det(g'_{ab}) = \det\left(\frac{\partial x^i}{\partial y^a}\right)^2 g_{ij}. \tag{31}$$

$$dVol = \sqrt{\det(g_{ij})} dx^1 dx^2 dx^3 = \sqrt{\det(g'_{ab})} dy^1 dy^2 dy^3. \tag{32}$$

$$h_{ab} = \frac{\partial x^i}{\partial w^a} g_{ij} \frac{\partial x^j}{\partial w^b} = \frac{\partial y^a}{\partial w^a} g'_{ab} \frac{\partial y^b}{\partial w^b}. \tag{33}$$

$$dA = \sqrt{\det(h_{ab})} dw^1 dw^2 \qquad \text{is unchanged.} \tag{34}$$

The invariants $S$ and $V$ have now been constructed for an arbitrary, bounded, smooth, manifold-with-boundary embedded in Euclidean 3-space.

**6.2.  Definition of the radial parameter, $R$, in terms of extrinsic curvature of Riemann.**

Let $u$ be a tangent vector at point $p$ in $\P B$ and let $g:[0,1] \to \P B$ be a curve with $\gamma(0) = p$. In general, if $u$ is translated along the curve $g$ using parallel translation of the embedding 3-space, the translated vector will no longer be tangent to $\P B$. An acceleration $a$ is needed at each point of $g$ in order to keep the translated vector parallel to ¶*B* as it is translated along ***g***. This acceleration can always be chosen normal to ¶*B* because parallel translation in 3-space and parallel translation of tangents to ¶*B* both preserve lengths. The extrinsic curvature of the surface ¶*B* is the length of the acceleration vector needed to make a tangent to ¶*B* remain tangent to ¶*B* during translation. At each point of ¶*B*, the extrinsic curvature depends on two vectors tangent to ¶*B*: the translated vector (***u*** in this discussion); and the direction of motion (*d**g**/dt* in the discussion above). Therefore the extrinsic curvature is a two-index tensor, or matrix, denoted $EK_{\alpha\beta}$. $EK_{ab}$ describes how the surface ¶*B* sits in 3-space; and it describes how the differentiation of vectors on ¶*B*, treated as vectors in 3-space, differs from such treated as vectors on ¶*B*. If the coordinates $w^a$ are orthonormal at a point on ¶*B*, then the eigenvalues $1/R_1$ and $1/R_2$ of *EK* are the principle curvatures of ¶*B* at that point, and the respective eigenvectors of *EK* are the directions of the principle curvature of the surface ¶*B* at that point.

Let $^3\nabla$ denote covariant differentiation in the Euclidean 3-space; $^2\nabla$ denote covariant differentiation induced from 3-space on the manifold ¶*B*; and ***n*** denote the unit normal vector field to ¶*B*, taking the normal vector field of either sign. For a pair of vectors ***u*** and ***v***, the extrinsic curvature EK is given by:

$$\mathbf{u} \cdot \mathbf{EK} \cdot \mathbf{v} = \mathbf{n} \cdot \left( {}^3\nabla_{\mathbf{u}} \mathbf{v} - {}^2\nabla_{\mathbf{u}} \mathbf{v} \right). \tag{35}$$

which is, in index notation,

$$\mathbf{u}^a \mathbf{EK}_{ab} \mathbf{v}^b = \mathbf{n}^i g_{ij} \mathbf{u}^k \left( {}^3\nabla_k \mathbf{v}^j - {}^2\nabla_k \mathbf{v}^j \right). \tag{36}$$

Let the surface be specified by the functions $x^i(w^a)$ ($i = 1,2,3$; $a = 1,2$), and let ***n*** be the normal vector field on the surface. Then

$$\mathbf{EK}_{ab} = \frac{\partial x^i}{\partial u^a \partial u^b} g_{ij} n^i = -\frac{\partial x^i}{\partial u^a} g_{ij} \frac{\partial n^j}{\partial u^b}. \tag{37}$$

Simple computational expressions for the extrinsic curvature with derivations are given in O'Neill[35]. Computational expressions for the extrinsic curvature can be found in standard references on differential geometry.[33-35]

The coordinate-invariant expression for the mean extrinsic curvature at a point *w* on ¶*B* is:

$$\frac{1}{2} EK_{ab}(w) h^{ab}(w), \tag{38}$$

where $h^{ab}(w)$ is the inverse matrix of $h_{ab}(w)$. In terms of the principal radii of curvature, $R_1$ and $R_2$,

$$Ek_{ab} h^{ab} = \left( \frac{1}{R_1} + \frac{1}{R_2} \right). \tag{39}$$

The mean extrinsic curvature averaged over the surface ¶*B* is the integral of the mean curvature with respect to the area form:

$$M = \int_{\partial B} \frac{1}{2} Ek_{ab} h^{ab} \sqrt{\det(h_{ab})} dw^1 dw^2. \tag{40}$$

The radial parameter, *R*, is now defined as $R = M/(4\pi)$ in order to be consistent with the convention used by Kihara and Boublík.

The invariant *R* has now been constructed for an arbitrary, bounded, smooth, oriented manifold-with-boundary embedded in Euclidean 3-space. This definition yields a unique result for surfaces which are both convex and non-convex, and also for surfaces which are multiply-connected. The surface must be oriented in order that *EK* is defined globally on the surface because the definition of *EK* depends on a choice of normal vector field on the surface. (The embedding 3-space is assumed to be oriented; and the right-hand rule and an orientation on the surface into a normal vector field have been used.)

In his 1953 paper[20] (page 840), Kihara wrote of the average of his support function, $H$, "… $4p<H>_{Ave}$ is equal to the mean curvature integrated over the whole surface of the convex body…" His equation (7.6) indicates that his parameter $M$ equals the general integral for the mean extrinsic curvature. However, Kihara's terminology makes no distinction between the extrinsic curvature (which is a property of the embedding and not of the metric surface), and the intrinsic Riemann curvature (which is a property of the metric surface itself). Kihara's method of computing $M$ depends on the convexity of the body enclosed by the surface, or, more precisely, on the star-shaped property that there exists a point inside the enclosed volume which has a clear line of sight to all points on the boundary surface. Kihara makes no mention of extending the definition of $M$ to non-convex (or non-star-shaped) bodies. Such computations are best performed using tensor calculus that is beyond the scope of the vector calculus used by Kihara. The computation of the mean curvature curvatures for many figures is given in the next section using tensor calculus. The computations were executed using the *Macsyma* software package, which has extensive symbolic integration capabilities, and which introduced the automatic computation of extrinsic curvature around 1996.

## 7. Explicit development of the geometric parameters $R$, $S$, and $V$ for convex bodies using metric differential geometry.

In this section, the geometric parameters, $R$, $S$, and $V$, are developed explicitly for the most common convex hard bodies. For this development, the symbolic mathematics program *Macsyma* has been used to compute the mean extrinsic curvature, $R$, and, for some figures, the surface area or enclosed volume. The geometric parameters $R$, $S$, and $V$ are defined as follows: $S$ = surface area, $R = 1/(4\pi) \times$(mean extrinsic curvature) of that surface, $V$ = volume subtended by $S$.

### 7.1 Spheriods: Spheres; ellipsoids of revolution; spherocylinders.

*Sphere*

For a sphere of radius $B$, the geometric parameters are:

$$\left. \begin{array}{l} R = B \\ S = 4pB^2 \\ V = \dfrac{4p}{3}B^3 \end{array} \right\}. \tag{41}$$

*Spherical section*

For a spherical sector of a sphere of radius $B$ with polar angle $\theta$, from which has been excised a spherical sector with angle $q$ between the polar axis and the edge of the excised sector. The geometric parameters are:

$$\left. \begin{array}{l} R = (1 - \cos q) B/2 \\ S = 2p(1 - \cos q) B^2 \\ V = p(1 - \cos q)^2 (2 + \cos q) B^3/2 \end{array} \right\}. \tag{42}$$

*Oblate spheroid*

An oblate spheroid has three principal radii A, B and C where A = B = C. The geometric parameters are:

$$R = \frac{A^2\sqrt{A^2-C^2}\arctan\left(\frac{\sqrt{A^2-C^2}}{C}\right) - C^3 + A^2C}{2(A^2-C^2)}$$

$$S = 2\pi A\left(\frac{c^2\log\left(\frac{\sqrt{A^2-C^2}}{C} + \frac{\sqrt{A^2-C^2}+1}{C}\right)}{\sqrt{A^2-C^2}} + A\right) \tag{43}$$

$$V = \frac{4}{3}\pi ABC$$

*Prolate spheroid*

A prolate spheroid has three principal radii $A$, $B$ and $C$ where $A = B = C$. The geometric parameters are:

$$R = \frac{A}{2}$$

$$S = 2\pi C \frac{A^2 \arcsin\left(\frac{\sqrt{A^2-C^2}}{A}\right)}{\sqrt{A^2-C^2}} \tag{44}$$

$$V = \frac{4}{3}\pi ABC$$

The result for the surface area $S$ is taken from the *CRC Math Tables*, page 318.[36]

## 7.2 Cylinders.

*Right Circular cylinder*

Consider a cylinder of radius $A$ and axial length $L$. The geometric parameters are:

$$R = \frac{L}{4}$$

$$S = 2\pi AL \tag{45}$$

$$V = \pi A^2 L$$

*Circular cylinder with parabolic radius*

Let the axial length of the cylinder be $L$, and the radius of the cylinder be $p + qz^2$, where $z$ is the axial coordinate. Then

$$R = -\frac{(4pq-1)\arctan(qL) - qL}{8q}$$

$$S = \frac{\pi^2\left((16pq-1)\operatorname{arcsinh}(qL) + \sqrt{q^2L^2+1}\left(2q^3L^3 + (16pq^2+q)L\right)\right)}{8q^2} \tag{46}$$

$$V = \frac{\pi(3q^2L^5 + 40pqL + 240p^2L)}{240}$$

*Circular cylinder with quartic radius*

Let the axial length of the cylinder be L, and the radius of the cylinder be $p + qz^2 + rz^4$, where $z$ is the axial coordinate. Then

$$R = \frac{L\begin{pmatrix} 420pr^3L^8 - 5040pq^3r^2L^8 + 665q^2r^2L^8 + 5040pq^6rL^8 - 1050q^5rL^8 - 1120pq^9L^8 + 280q^8L^8 \\ +2520pqr^2L^6 - 135r^2L^6 - 5040pq^4rL^6 + 990q^3rL^6 + 1440pq^7L^6 - 360q^6L^6 + 5040pq^2rL^4 \\ -882qrL^4 - 2016pq^5L^4 + 504q^4L^4 - 5040prL^2 + 3360pq^3L^2 - 840q^2L^2 - 10080pq + 5040 \end{pmatrix}}{20160}$$

$$S = \frac{-p^2L\begin{pmatrix} 840pq^2r^2L^8 - 280qr^2L^8 - 840pq^5rL^8 + 315q^4rL^8 + 175pq^8L^8 - 70q^7L^8 \\ -720pr^2L^6 + 1440pq^3rL^6 - 900q^2rL^6 - 360pq^6L^6 + 180q^5L^6 - 4032pqrL^4 \\ -504rL^4 + 1008pq^4L^4 - 1008q^3L^4 - 6720pq^2L^2 - 3360qL^2 - 40320p \end{pmatrix}}{10080}$$

$$V = \frac{pL(35r^2L^8 + 360qrL^6 + 2016prL^4 + 1008q^2L^4 + 13440pqL^2 + 80640p^2)}{80640}$$

(47)

*Circular cylinder with inverted circular radius*

Let the axial length of the cylinder be $L$, and the radius of the cylinder be $(q + p - \sqrt{(q^2 - z^2)})$, where $z$ is the axial coordinate of the cylinder. In an $r$–$z$ cross-section, the radius of this cylinder is formed by a circle of radius $q$ whose point of closest approach to the axis of the cylinder is $p$. In order to evaluate the integral for the mean extrinsic curvature, a Taylor series of the above expression to the eighth order is used. All results quoted for the inverted-circle cylinder for $R$ and $S$ after this point use the eighth-order Taylor approximation for the radius. The results for the volume are exact.

$$R = \frac{-(q+p)(1225L^9 + 7200q^2L^7 + 48384q^4L^5 + 4300080q^6L^3) + 10321920(q-p)q^8L}{41287680q^9}$$

$$S = \frac{p(q+p)(1225L^9 + 7200q^2L^7 + 48384q^4L^5 + 4300080q^6L^3) + p10321920pq^8L}{5160960q^8}$$

(48)

$$V = \frac{p\left(12p^2L - L^3 + 6(q+p)\left(4qL - 4q^2\arcsin(L/2q) - L\sqrt{4q^2 - L^2}\right)\right)}{12}$$

### 7.3 Shapes with sharp edges.

*The general case*

Consider two flat surfaces which meet at an angle **q** along a straight edge of length *L*. Let the edge be rounded with radius *r*. The principal curvature transverse to the edge is $1/r$, and the principal curvature along the edge is zero. Therefore the mean extrinsic curvature is $1/(2r) = \frac{1}{2}(1/r + 0)$ on the curved sector of angle **q**, and zero on the flat surfaces. The area of the curved sector is $r\mathbf{q}L$, so the integral of the mean extrinsic curvature over the entire surface is ½ **q** L. Dividing by 4**p** in order to be consistent with the convention used by Kihara, the geometric shape parameter *R* is:

$$R = \frac{\mathbf{q}L}{8\mathbf{p}}.$$

(49)

This result is independent of the radius of curvature *r*, so it holds for a sharp edge with exterior angle **q**.

More generally, consider an edge with a sharp angle (or rounded with radius *r*) with parameter *z* along the length of the edge. Let the exterior angle at which the two surfaces meet be $q(z)$. Then the contribution of the edge to the geometric parameter *R* of the surface is:

$$\Delta R = \int \frac{q(z)}{8p} dz. \tag{50}$$

In particular, if $q$ is constant along an edge of length *L*, then the contribution of the edge to the parameter *R* is:

$$\Delta R = \frac{qL}{8p}. \tag{51}$$

These last two results also hold for a sharp edge that is curved along its axis, because the principal curvature along the edge is finite, and the domain of integration as a two-dimensional surface is infinitely narrow. Some specific examples of figures containing sharp edges are now computed.

*Rectangular Solid*

Consider a rectangular solid with edge lengths A, B and C. The geometric parameters of this figure are:

$$\left.\begin{array}{l} R = \dfrac{(A+B+C)}{4} \\ S = 2(AB+BC+CA) \\ V = ABC \end{array}\right\} . \tag{52}$$

*Infinitely Thin Planar Figure*

Viewed as a 3-D figure, a planar figure has an exterior angle of 2p along its boundary. The geometric parameters are:

$$\left.\begin{array}{l} R = \dfrac{\text{circumference}}{4} \\ S = 2 \times \text{Area-of-one-side} \\ V = 0 \end{array}\right\} . \tag{53}$$

More generally, for any 2-D surface with boundary, the above result describes the contribution of the boundary to *R*.

Table 1 Geometric parameters for typical convex hard-body particles, as determined by differential geometry. For spheroidal and ellipsoidal hard bodies, $s$ is the diameter; for cylinders and spherocylinders, $g$ is the length-to-breadth ratio; for ellipsoids of revolution, $l$ is the ratio of the major-to-minor axes.

| hard body | parameters | *R* | *S* | *V* |
|---|---|---|---|---|
| convex hard bodies ||||| 
| sphere | σ | σ/2 | πσ² | 1/6πσ³ |
| cylinder | γ, σ | 1/4(γ+π/2) σ | π(γ+1/2) σ² | 1/4πγσ³ |
| prolate spherocylinder | γ, σ | 1/4(γ+1) σ | πγσ² | 1/12π(3γ-1) σ³ |
| oblate spherocylinder | ϕ(= γ-1), σ | 1/2(πϕ/4+1)σ | 1/2π(ϕ²+πϕ+2) σ² | 1/24π(6ϕ²+3πϕ+4)σ³ |
| prolate ellipsoid of revolution | λ>1, σ | $\dfrac{1}{4}\left[1+\dfrac{\ln\left(l+\sqrt{l^2-1}\right)}{\sqrt{l^2-1}}\right]s$ | $\dfrac{1}{2}p\left[1+\dfrac{l^2\arccos\left(l^{-1}\right)}{\sqrt{l^2-1}}\right]s^2$ | $p/6ls^3$ |
| oblate ellipsoid of revolution | λ<1, σ | $\frac{1}{4}\left[\boldsymbol{l}+\frac{\arccos(\boldsymbol{l})}{\sqrt{1-\boldsymbol{l}^2}}\right]\boldsymbol{s}$ | $\frac{1}{2}p\left[1+\frac{\boldsymbol{l}^2}{\sqrt{1-\boldsymbol{l}^2}}\ln\left(\frac{1-\sqrt{1-\boldsymbol{l}^2}}{\boldsymbol{l}}\right)\right]\boldsymbol{s}^2$ | $p/6\boldsymbol{l}\boldsymbol{s}^3$ |
| regular parallelepiped | a, b, c | 3(a + b + c)/4 | 2(ab + bc + ac) | abc |

### 7.3 Conditions Ensuring that Kihara's *R* is the Mean Extrinsic Curvature.

That the mean curvature of Riemann gives the averaged center-to-center half-distance of two convex hard-body particles as one rolls upon the entire surface of the other raises immediately the question, to what extent does this equality hold for non-convex hard bodies. Here follow certain conditions on the principal extrinsic curvatures of ¶*b* which ensure that the mean extrinsic curvature yields the same result as Kihara's construction of *R* in terms of rolling molecules over one another.

Assume that the maximum magnitude of a principal radius of curvature on any concave patch of ¶*b* is less than the minimum magnitude of a principal radius of curvature on any convex patch of ¶*b*, that is,

$$\left|R_{\max}\left(\partial\boldsymbol{b}_{\text{patch}}\right)^{\text{concave}}\right| \quad < \quad \left|R_{\min}\left(\partial\boldsymbol{b}_{\text{patch}}\right)^{\text{convex}}\right|. \tag{54}$$

When this condition, (54), is satisfied, then any part of ¶*b* fits tangentially to any other part of ¶*b*, at least locally. Such a local curvature condition cannot guarantee that tangency of arbitrary points can be arranged, because of possible interference of distance parts of the molecules. If this curvature condition is satisfied, the integral of the mean extrinsic curvature over the entire surface is identical to Kihara's parameter *R*, except for the possibility of interference if distant parts of the molecules in Kihara's construction.

If the condition (54) is not satisfied, then it may be necessary to exclude (or otherwise adjust for) certain concave patches of ¶*b* from the surface integral defining the mean extrinsic curvature, in order that the result is identical to the result of Kihara's construction in terms of molecules rolling over one another.

### **8.** Development of the geometric parameters *R*, *S*, and *V* for non-convex hard-body particles using metric differential geometry.

In this section are developed the geometric parameters, *R*, *S*, and *V*, for hard bodies which are explicitly non-convex. In the cases which follow, the geometric parameters are developed partly as a formal exegesis and partly for reference. Most of the particular hard bodies described in this section are *not* those which are applicable for a statistical mechanical argument, for their respective surfaces do not represent the surfaces of exclusion, $S_0$, into which no part of another particle can intrude, as used in equation **(4)**.

## *N* tangent spheres.

Consider a figure consisting of *N* spheres that are tangent to one another. Whether the spheres are arranged in a linear chain or a branched or ring structure has no effect on the geometric parameters *R*, *S*, and *V*. The points of tangency contribute nothing to either the surface area or the volume of the figure. They also contribute nothing to the curvature parameter *R*, because each is a sharp edges of exterior angle $\pi$ and length zero. Therefore the geometric parameters *R*, *S*, and *V*, for a figure consisting of *N* tangent spheres is just *N* times the parameters for one sphere.

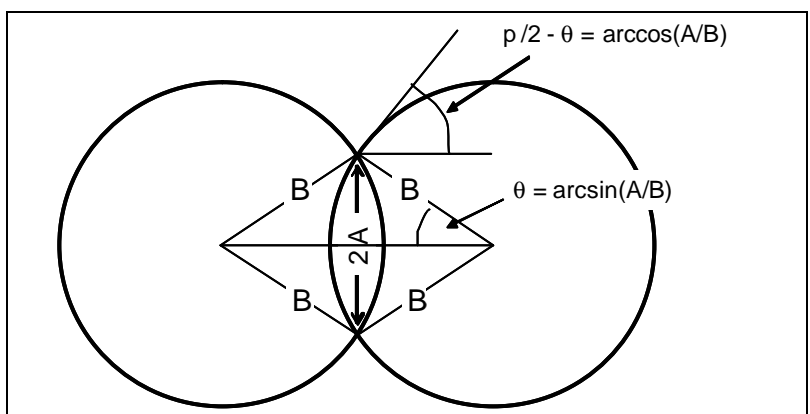

**Fig. 3. Overlapping spheres.**

Tangent spheres are not good models of chemical bonds nor even of clustered particles, for their surfaces do not faithfully represent the regions of space from which other particles are excluded. Nonetheless, the geometric figures described below include regions that may be considered to include bonds between atoms; and their geometric parameters *R*, *S*, and *V* are computed below and will be compared to the parameters appropriate for physical molecules.

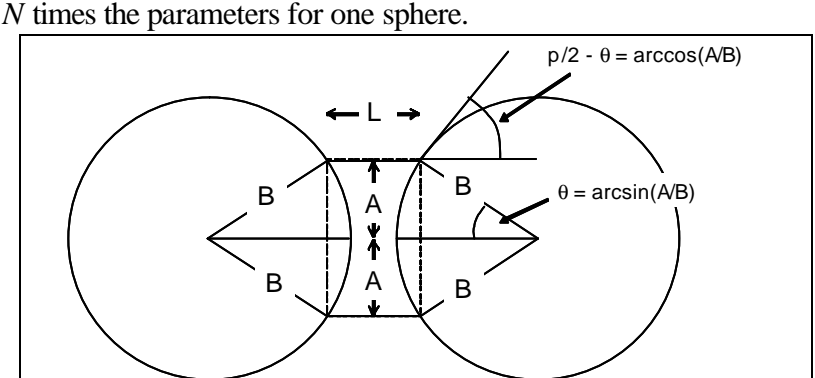

**Fig. 4. Chain of spheres with cylindrical bonds.**

## A linear chain of overlapping spheres.

Consider a linear chain of *N* overlapping spheres of radius *B*. The two spheres each have one region of overlap with another sphere. The *N*-2 spheres that are not on the ends have two sectors of overlap with other spheres. Let $q$ = the half-angle of the circle of overlap between the spheres.

Each seam where the spherical surfaces meet with exterior angle of $2q - \pi$ contributes to *R*. Each seam contributes [(exterior angle)×circumference]/($8\pi$) = $(2q-\pi)\times A/4$. The geometric parameters for a chain of *N* overlapping spheres are:

$$\left. \begin{aligned} R &= \left(2\pi(n-1)(2q-\pi)\sin q + (n-1)\cos q + 1\right)B \\ S &= 4\pi\left((n-1)\cos q + 1\right)B^2 \\ V &= \frac{2\pi}{3}\left(2-(n-1)\cos q\left(\cos^2 q - 3\right)\right)B^3 \end{aligned} \right\}. \qquad (55)$$

Note that $A = 0$, if and only if $q = 0$, if and only if the spheres are tangent.

## Linear chain of spheres with cylindrical bonds.

Consider a linear chain consisting of *N* spheres of radius *B* that are connected by cylinders of radius *A*, where $A = B$. Let the axial length of each cylinder be *L*. The conditions for overlapping, tangent, or disjoint spheres are

$$2B(1-\cos q) = 2\left(B - \sqrt{B^2 - A^2}\right) \quad > \quad L \quad \text{overlapping spheres}$$
$$2B(1-\cos q) = 2\left(B - \sqrt{B^2 - A^2}\right) \quad = \quad L \quad \text{tangent spheres} \quad \quad (56)$$
$$2B(1-\cos q) = 2\left(B - \sqrt{B^2 - A^2}\right) \quad < \quad L \quad \text{disjoint spheres}$$

where $q = \arcsin(A/B)$. The geometric parameters are:

$$R = \frac{(n-1)L + 4(n-1)\sqrt{B^2 - A^2} + 4B + 4p(n-1)A(2\arcsin(A/B) - p)}{4}$$
$$S = 2p\left((n-1)AL + 2B\left((n-1)\sqrt{B^2 - A^2} + B\right)\right) \quad (57)$$
$$V = p(n-1)A^2 L + \frac{4p}{3}B^3 + \frac{2p(n-1)\sqrt{B^2 - A^2}(2B^2 + A^2)}{3}$$

**Linear chain of tangent spheres with inverted-circle cylindrical bonds:**

This case is identical to the previous one, except that the spheres are tangent, and straight cylinders are replaced by cylinders whose radii have the shape of inverted circles. An example of this case is shown in Fig. 5 for three touching spheres, arranged as a linear collineation. Recall that results for $R$ and $S$ employ an eighth-order Taylor series approximation for the radius of the cylinder, so they contain either polynomials in $L$ and, when $L$ is solved for, large integers. Floating point approximations for the geometric parameters R, S, and V for this special case are:

$$R = (0.54656n + 0.453544)B$$
$$S = (11.3961n + 1.17024)B^2 \quad (58)$$
$$V = (4.77376n - 0.58497)B^3$$

Unlike the previous models of overlapping spheres and chains with cylindrical bonds, this particular model *is* particularly relevant for the analysis of linear clusters of molecules, for its surface plainly *does* represent faithfully that into which no part of another sphere can penetrate. This model represents the linear clusters of spherical molecules which participate in the gas-liquid phase transition, as will be shown in an article to follow.

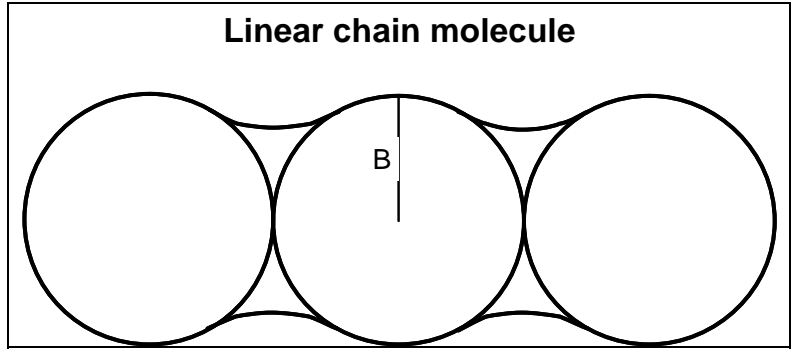

**Fig. 5** The effective surface of exclusion, which subtends the volume into which no part of another sphere may penetrate, for a linear collineation of three touching spheres.

### 9. Experimental validation of the use of the geometric parameters developed by differential geometry for the statistical mechanical description of real molecules.

In the previous sections, it has been shown that often an infinite number of hard-body molecules of quite different geometric parameters $R$, $S$, and $V$ can manifest the same Boublík coefficient $a$, and that furthermore, if they possess the same intrinsic volume of exclusion, $V$, they will have identical thermodynamic proper-ties as pure components. In this section, this result is applied and shown to hold true for a set of real molecules, the normal alkanes.

In Fig. 6, is shown a ball-&-stick representation of the real molecule n-octane, a typical member of the class of normal alkane compounds. N-octane is plainly not a convex molecule, as also are none of the n-alkanes. However, the n-alkanes may be considered to possess a statistically averaged axis of rotation along its length. Because the n-alkanes have the properties of chains of $CH_2$ radicals, they possess also an average length-to-breadth ratio. Therefore, one may reasonably model the n-alkane molecules effectively as prolate spherocylinders, and calculate their respective Boublík coefficients of asphericity, $\boldsymbol{a} = RS/3\boldsymbol{u}$ using that model.

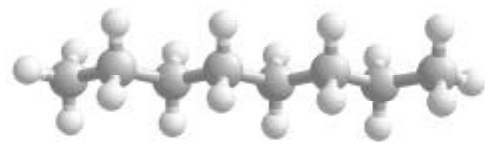

**Fig. 6 Model of the typical normal alkane molecule, n-octane.**

The thermodynamic properties of the normal alkanes have been analyzed using the formalism of the Simplified Perturbed Hard-Chain Theory[37-41] [SPHCT]. The SPHCT uses first-principles statistical mechanics arguments to develop the following equation of state:

$$\boldsymbol{b}\, p^{\text{SPHCT}} \;=\; \boldsymbol{b}\left(p^{\text{ref}} + p^{\text{vdW}}\right) = \boldsymbol{r}\left[\left(1 + c^{\text{SPHCT}}\frac{2\boldsymbol{h}'(2-\boldsymbol{h}')}{(1-\boldsymbol{h}')^3}\right) - c^{\text{SPHCT}} Z_m \ln\left(1+\boldsymbol{h}'Y/\boldsymbol{t}\right)\right] \tag{59}$$

In equation (59), the SPHCT packing fraction, $\boldsymbol{h}' = N\boldsymbol{t}V^*/V$, where $\boldsymbol{t} = \pi\sqrt{2}/6 \approx 0.7405$ is the close-packing parameter, the ratio of the total volume of close-packed spheres to their enclosing volume. (The factor $\boldsymbol{t}$ in equation (59) always appears multiplying the SPHCT volume parameter, $V^*$, by convention. Originally, the singularity of the scaled particle theory factor in the equation of state was thought to be "incorrect," and the factor $\boldsymbol{t}$ was introduced *ad hoc* to "correct" it. The result is only that the true, hard-core parameter for the SPHCT equation is $\boldsymbol{t}V^*$, and its volume parameters, $V^*$, are always conventionally stated oversized.) The first two terms on the right side of equation (59) represent the contribution to the pressure from the hard-core reference system; the last term represents the contribution of the long-range, attractive, van der Waals component of the intermolecular potential, in which $Y = \exp(T^*/2T) - 1$, where $T^*$ is a characteristic temperature of the system.

In equation (59), the parameter, $c^{\text{SPHCT}}$, multiplying the contributions to the pressure both from the hard-core and the van der Waals components of the intermolecular potential, is known as the Prigogine "*c*-factor." Prigogine originally argued[42, 43] that the *c*-factor should account for a modification of the dimensions of the partition function of the system attributable to the interaction of certain of the of molecular vibrations and rotations with the translational degrees of freedom. Prigogine's original arguments do not stand up to scrutiny; and it has been shown[11] that his "*c*-factor" is, in fact, a shape-factor that accounts somewhat, at low densities, for the effects of molecular asphericity and approximates the Boublík coefficient of asphericity, $(1+3\boldsymbol{a})/4$, in the low-density regime.

When the Prigogine *c*-factor is recognized as an approximation to the Boublík coefficient of asphericity, **a**, the product of the Prigogine *c*-factor and the SPHCT hard-core volume, $tV^*$, can be related directly to the factor of the function of the Boublík coefficient and the geometric volume parameter, $u^B$, through the second virial coefficient. The hard-core component of the second virial coefficient for the *hard-core reference system, as developed by the SPHCT is:

$$B_2^{SPHCT} = 4c^{SPHCT} t V^{*SPHCT}. \tag{60}$$

The second virial coefficient, as given by scaled particle theory, of a hard-body gas is:

$$B_2^{SPT} = u^G \left(1 + 3a^G\right). \tag{61}$$

in which $u^G$ and $a^G = R^G S^G / 3u^G$ are, respectively, the volume parameter determined by differential geometry and the Boublík coefficient of asphericity. For prolate spherocylinders, the Boublík coefficient of asphericity, **a**, is:

$$a^G = \frac{g(g+1)}{(3g-1)}, \tag{62}$$

where **g** is the ratio of the length to the diameter of the spherocylinders, which for n-alkanes may be taken as the carbon number.

Equations (60) and (61) taken together allow the geometric parameters determined by differential geometry to be related directly to experimentally measured the Prigogine *c*-factors through the identity of their second virial coefficients.

$$\boxed{c^{SPHCT} = \frac{1}{4t}\left(1 + 3a^B\right)}. \tag{63}$$

This property is demonstrated in Fig. 7 where are plotted the values of the Prigogine *c*-factor for the first ten n-alkanes, fitted by van Pelt *et al.*[44] to the experimental values of the pressure-density curves, and also the calculated Boublík asphericity parameters, **a**. In Fig. 7, the Prigogine *c*-factors are plotted as scatter points in white squares; the Boublík asphericity parameters are plotted as black circles and a dotted trace.

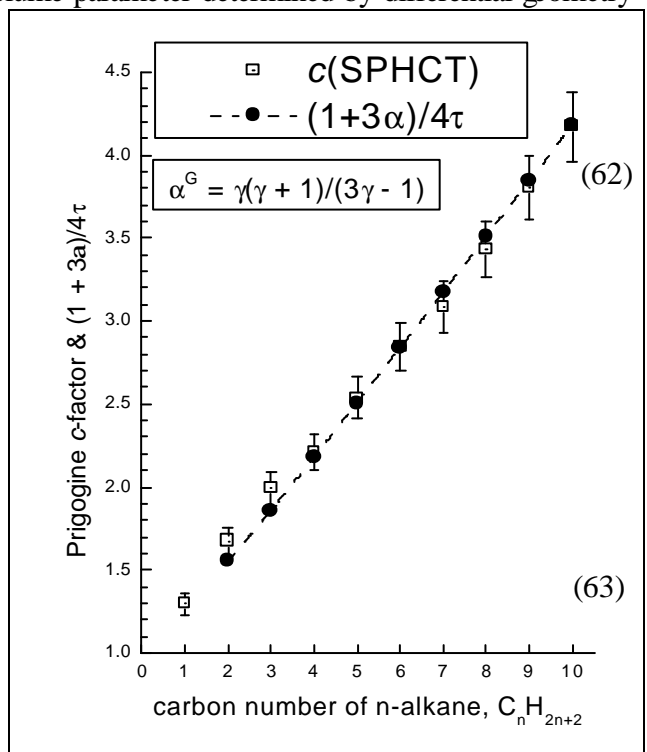

**Fig. 7 Comparison of Prigogine *c*-factor fitted by van Pelt *et al.*[44] for n-alkanes, with the n-alkanes modeled as spherocylinders.**

The agreement shown in Fig. 7 is obvious. It cannot be emphasized too strongly that the white squares, which represent the respective Prigogine *c*-factors, were obtained solely by fitting the SPHCT equation of state to the experimental pressure-density curves. There is *no* underlying theoretical reasoning behind the values of those white squares; they represent solely experimental data fitted to the SPHCT equation of state, (59). Similarly it cannot be emphasized too strongly also that the black circles and the dotted line connecting them, which represent the values of $(1+3a)/4t$, were obtained solely from equation (62), with *no* experimental input. There have been *no* added fitting parameters used.

## 10. Discussion.

The geometric parameters of a hard body, *S*, *V*, and *R*, which describe respectively the surface and volume of exclusion into which no part of another particle can be injected, and the mean half-distance of one particle as it rolls upon another in all possible orientations, have been extended using metric differential geometry. These radial parameters have been demonstrated to be the same parameters *R*, *S*, and *V* developed by Kihara for convex hard-bodies, and have been extended for non-convex and multiply-connected bodies (like tori or donuts). This extension has been based upon the extrinsic curvature of ($n$-1)-dimensional surfaces embedded in Euclidean $n$-space. We present *R*, *S* and *V* for a range of convex and non-convex shapes, including chains of spheres in which molecular bonds are represented by sharp seams, or straight cylinders, or by "inverted-circle" cylinders.

### 10.1 The thermodynamic ambiguity of the geometric properties of single-component, hard-body systems.

The statistical thermodynamic description of a single-component, hard-body system in terms of its geometric parameters of *R*, *S*, and *V*, is *not* unique. As set forth in the Boublík equation of state, (18), the thermodynamic properties of the system are described completely by is volume of exclusion, *V*, and its respective Boublík parameter of asphericity, **a** = $RS/3V$. For given values of the geometric parameters *V* and **a**, there exist infinitely many different geometric shapes. For example, a prolate spherocylinder with aspect ratio (length-to-breadth) **g** = 6 and volume *V* = 45, will have the cross-section **s** = 2.1624 and Boublík parameter **a** = 2.470. There exist also infinitely many parallelepipeds with the same volume and Boublík parameter. This property is demonstrated graphically in Fig. 8 in which every point on the solid line represents a parallelepiped with the same parameters *V* and **a**. The solid trace in Fig. 8 represents the ratio of the surface parameter of a parallelepiped to the length of its longest side, for a chosen typical magnitude of one of the other sides. In addition to the parallelepipeds with the maximum and minimum values for this ratio, there plainly exists infinitely other parallelepipeds with the same parameters *V* and **a**. Similarly there exist also infinitely many ellipsoids with the same parameters, and also infinitely many irregular hard-bodies.

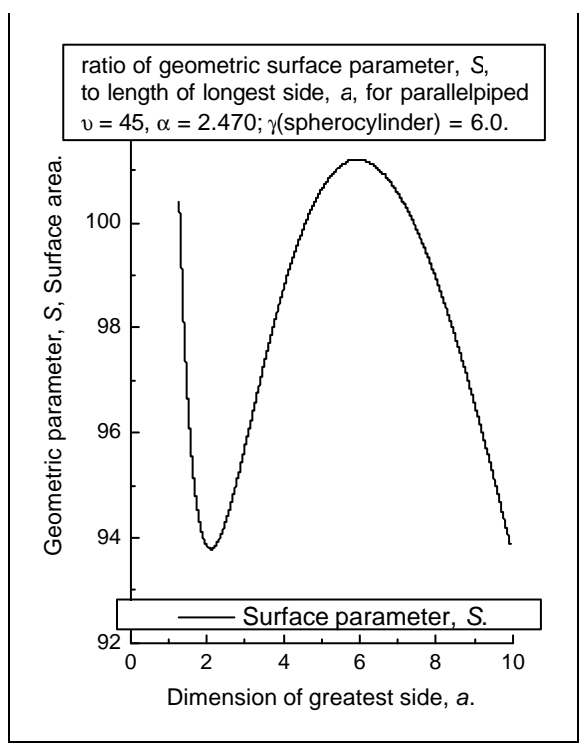

**Fig. 8 Ratio of geometric surface parameter, *S* to length of longest side, a, for a parallelepiped with the identical volume, *u*, and Boublík coefficient of asphericity, *a*, as a prolate spherocylinder of aspect ratio, *g*= 6.**

Thus, although the geometric properties of a hard-body determine its statistical thermodynamic properties, the inverse statement is not true. The volume parameter *V* uniquely determines the packing-fraction **h** = **r***V*. However, that parameter together with the surface parameter and its mean exterior curvature enter the equation of state only in combination through the Boublík parameter of asphericity, **a**. Therefore, the geometric parameters of the hard-body system are not one-to-one with its thermodynamic properties, but one-to-many.

In this respect, the geometric description of the hard-body system through is parameters *R*, *S*, and *V* determines the statistical mechanical behavior of a system analogously as the parameters describing the inertial ellipsoid of a rigid body determines its mechanical behavior: Just as a rigid body shaped as an ellipsoid will have exactly the dynamical properties of that body, so also will infinitely many other, - and, typically, very irregular, - rigid bodies which possess the same ellipsoid of inertia.

### 10.2 The thermodynamic specificity of the geometric properties of multi-component, hard-body systems.

The thermodynamic ambiguity of a single-component, hard-body system is largely lifted by the presence of another component. The Boublík equation, (18), has the same form for multi- and single-component systems. However, the Boublík coefficient of asphericity for multi-component systems involves products and quotients of weighted sums of the individual geometric parameters $R_i$, $S_i$, and $V_i$, and is sensitive to differences in shape. The differences in the individual geometric parameters $R_i$, $S_i$, and $V_i$, induce an excess volume on the mixed system, and thereby changes in the Gibbs free enthalpy. At increased densities, the change in the Gibbs free enthalpy typically induces phase separation.

This property of the excess volume induced by geometric differences of hard-body particles has been demonstrated for the most demanding case involving subtle geometric differences: the phase separation of optical enantiomers. For optical enantiomers, the respective magnitudes of the geometric parameters $R_i$, $S_i$, and $V_i$ are identical, and differ only in their respective chirality.[12] A formal development of the general mathematical technique for projecting out the respective chiral and achiral components of the geometric parameters *R*, *S*, and *V* is given in Appendix C.

### 10.3 The thermodynamic description of a hard-body system at high density

The rigorous development by differential geometry of the geometric parameters of the hard-body reference system admits accurate thermodynamic description of the system not only at both high and low densities. The thermodynamic analysis is no longer restricted to convex hard-body particles in the low-density regime and also can describe the system when the available volume becomes multiply connected in the high-density regime.

In the low-density regime, the available volume of the fluid, $V_0$, and the boundary of the space determined by that volume, $S_0$, are determined from the sums of the respective individual molecular geometric parameters, e.g., for a hard-sphere fluid, $V_0 = (V - N4/3\pi\sigma^3)$, $S_0 = N4\pi\sigma^2$. In the low-density regime, the ratio $S_0/V_0$ plainly increases with *N*.

In the high-density regime, the description of the available volume, $V_0$, and its surface, $S_0$, becomes more complicated. At high density, the available volume into which another particle can be injected becomes multiply-connected, and the roles of the "free" and "excluded" volumes become interchanged. At high density the available volume becomes distributed among "holes" in the matrix of the excluded volume. Reiss[17] and Hoover *et al*.[45] have named this high-density mode the "Swiss cheese model." In such high-density regime, as the density is increased the available volume becomes smaller with a consequential decrease in the surface to volume ratio $S_0/V_0$. This property suggests the presence of a two-phase region. Previously, this question could be addressed only qualitatively. Now, the multiply-connected available volume and its surface can be analyzed rigorously. In a paper to follow, the high-density phase stability of the hard-sphere fluid will be described explicitly.

**Appendix A. Derivation of the expression for the Gibbs free enthalpy of formation (chemical potential), $\mu(p,T;N)$, pressure, $p(V, T; N)$, and configuration factor, $Z(N)$, of the Canonical Partition Function.**

**A.1.1.   Derivation of expression (5) for the chemical potential, $\mu(p,T;N)$.**

To add a sphere of radius $b$ to the equilibrium configuration of the fluid, there must exist a cavity large enough to accommodate that sphere. The probability of finding such a cavity is:

$$P_0(r) = \frac{V_0(r)}{V} \tag{64}$$

The probability of finding such a cavity is given explicitly by the statistical mechanical theory of fluctuations in terms of the reversible work, $W(r)$, required to produce such a cavity:

$$P_0(r) = \exp(-\beta W(r)) \tag{65}$$

Equations (64) and (65) together give:

$$W(r)\beta = -\ln\left(\frac{V_0(r)}{V}\right) \tag{66}$$

When the radii of all particles are set equal to zero, the system reduces to an ideal gas, for which the chemical potential is:

$$\mu\beta = -\ln\left(\frac{V}{N\lambda^3}\right) \tag{67}$$

When the respective radii of the fluid, $a$, and of the cavity, $b$, are "scaled up" and set equal, the additional reversible work to be added to equation (67) is given by equation (66):

$$\mu\beta = \mu^{IG}\beta + W(a)\beta = -\ln\left(\frac{V}{N\lambda^3}\right) - \ln\left(\frac{V_0(a)}{V}\right) = \ln\left(\frac{N\lambda^3}{V_0}\right) \tag{68}$$

which is equation for the chemical potential, (5). One should note that when the chemical potential is written,

$$\mu = k_B T\left(-\ln\left(\frac{V}{N\lambda^3}\right) + \ln\left(\frac{V}{V_0}\right)\right) \tag{69}$$

the first term is always negative and the second always positive. This property is of fundamental significance for the phase stability of the hard-sphere system.

**A1.2.   Derivation of expression (4) for the system pressure, $p$.**

If there exists a spherical cavity of radius $r$ from which the centers of hard particles are excluded, the number density of such particle centers on that surface is $\rho G(r;\rho)$, where $G(r;\rho)$ is the conditional probability density for the center of a particle to be located at $r$ when it is known that there is no center of a particle within the cavity. Furthermore, when $r = (a+b)/2$ and the particles are in contact, a moment's consideration ascertains that

$$G(r;\rho) = G([a+b]/2;\rho) = g([a+b]/2;\rho), \tag{70}$$

where $g(r;\rho)$ is the standard two-body radial distribution function; for the case when $a = b$,

$$G(a;\rho) = g(a;\rho) \tag{71}$$

Because the transfer of momentum between a hard body and a hard cavity is purely kinetic, the equilibrium force per unit area on the surface of the cavity is:

$$f(r) = k_B T \rho G(r;\rho) \tag{72}$$

and the total force on an entire spherical surface is:

$$F(r) = 4\pi r^2 f(r) = k_B T 4\pi r^2 \rho G(r; \rho) \tag{73}$$

The reversible work required to increase the radius of the cavity is:

$$dW = F(r)dr = k_B T 4\pi r^2 \rho G(r; \rho)dr \tag{74}$$

such that

$$G(r; \rho) = \frac{1}{k_B T 4\pi r^2 \rho}\left(\frac{\partial W}{\partial r}\right)_T \tag{75}$$

From equation (66),

$$\left(\frac{\partial W}{\partial r}\right)_T = -k_B T \frac{1}{V}\left(\frac{\partial V_0}{\partial r}\right)_T \tag{76}$$

Remembering that $V_0(r; \rho)$ is the available free volume into which the center of another particle can be inserted, and that $S_0(r; \rho)$ is the area of the interface which bounds that volume and separates it from the excluded volume $(V - V_0)$, the incremental increase in the available free volume attributable to the incremental increase $dr$ of the dotted surface is:

$$dV_0 = S_0 dr \tag{77}$$

When equations (75), (76), and (77) are applied,

$$G(r; \rho) = \frac{1}{k_B T 4\pi r^2 \rho} \frac{S_0(r)}{V_0(r)} \tag{78}$$

When equation (71) is applied, the variable $r$ in equation (78) assumes the value $a$, and $G(r; \rho)$ becomes $g(a; \rho)$. Statistical mechanics gives the exact relationship for the pressure of a hard-sphere gas in terms of its two-body distribution function at the radius of contact[46]:

$$\frac{p\beta}{\rho} = 1 + \frac{2}{3}\pi \rho a^3 g(a; \rho) \tag{79}$$

When equation (71) is introduced into equation (78) for the contact value of $R_0 = a$, there obtains:

$$\frac{p\beta}{\rho} = 1 + \frac{R_0}{3}\frac{S_0}{V_0}, \tag{80}$$

which is the equation of state for the system pressure, (4).

### A1.3. Derivation of expression (6) for the configuration factor, $Z(N)$, of the canonical partition function.

The chemical potential of a particle in a single-component, $N$-body system is given directly by the Helmholtz free energy and the canonical partition function:

$$\mu = F_N - F_{N-1} = -k_B T \ln\left(\frac{Q_N}{Q_{N-1}}\right) = -k_B T \ln\left(\frac{1}{N\lambda^3}\frac{Z_N}{Z_{N-1}}\right) \tag{81}$$

where $F_N$ is the Helmholtz free energy, $Q_N$ the canonical partition function, and $Z_N$ the configuration integral, respectively, for a one-component system of $N$ particles. When equation (68) or (69) is applied directly to a system of 2 particles,

$$\mu = k_B T \ln\left(N\lambda^3 \frac{Z_1}{Z_2}\right) = k_B T \ln\left(\frac{2\lambda^3}{V_0(2)}\right). \tag{82}$$

Since $Z_1 = V$, then

$$Z_0(2) = VV_0(2) = V_0(1)V_0(2). \tag{83}$$

Having shown that equation (6) holds for $N = 1$, and $N = 2$, the process of mathematical induction can be applied whereby that equation is assumed to hold true for $N = k$ and thenafter demonstrated to hold identically for $N = k+1$. Thereby one has

$$Z_N = \prod_{n=0}^{N} V_0(n), \tag{84}$$

which is equation (6). Thus the configuration integral is shown to be the continued product of the available free volumes of all the $n$-particle systems from $n = 1$ to $n = N$.

## Appendix B.  Derivation of the exact solution for the equation of state of a of hard-sphere gas system by Scaled Particle Theory [SPT].

Scaled particle theory directly solves the thermodynamic problem of the hard-sphere gas. The connection between the scaled particle theory equation for the probability that there exists the center of a particle at distance $r$, the spherical volume of which is itself devoid of another particle, equation (9), and statistical thermodynamics comes through application of the pressure equation:

$$p\boldsymbol{b} = \boldsymbol{r}\left(1 + \frac{2}{3}\boldsymbol{\rho s}^3 \boldsymbol{r} G(\boldsymbol{s})\right). \tag{85}$$

Equation (85) is the *exact* equation from statistical mechanics for a fluid of hard spheres of radius $\boldsymbol{s}/2$.[46]

The probability that a particle, of diameter $\boldsymbol{s}$, might have its center somewhere within a sphere of radius less than $\boldsymbol{s}/2$ is simply $\boldsymbol{r}(4/3)\pi r^3$, because at most only one particle can exist within such sphere. Therefore, the probability that the sphere has no particle within it, $P(r)$, is $(1 - \boldsymbol{r}(4/3)\pi r^3)$. From the logarithmic derivative of $P(r)$ given by equation (8), the solution for the correlation function $G^-(r)$, valid for $r < \boldsymbol{s}/2$, obtains directly. For values of the radius greater than $\boldsymbol{s}/2$, the correlation function $G^+(r)$ is assumed to have the form $A + B/r + C/r$:

$$\left.\begin{array}{ll} G^-(r) = \dfrac{1}{\left(1 - \dfrac{4}{3}\boldsymbol{\rho} r^3 \boldsymbol{r}\right)} & r < \dfrac{\boldsymbol{s}}{2} \\[2ex] G^+(r) = A + \dfrac{B}{r} + \dfrac{C}{r^2} & r > \dfrac{\boldsymbol{s}}{2} \end{array}\right\}. \tag{86}$$

The model of the radial distribution function, $G^+(r)$, given by the second of equations (86) constitutes a formal assumption for the development of the equation of state for the hard-sphere gas by scaled particle theory. However, this assumption is not derived from scaled particle theory, and exists independently of it.

At the radius of the sphere, $r = \boldsymbol{s}/2$, both the forms of the correlation function and of its respective first derivatives must match. Furthermore, the kinetic pressure against any plane surface must correspond to the value of the correlation function $G^+(r)$ in the limit as $r$ increases without bound: $p = k_B T \boldsymbol{r} G^+(r)|_{r \to \infty}$.

$$\left.\begin{array}{lll} G^-\big|_{(s/2)} = G^+\big|_{(s/2)} & : & \dfrac{1}{\left(1 - \dfrac{1}{6}\boldsymbol{\rho s}^3 \boldsymbol{r}\right)} = A + \dfrac{2B}{\boldsymbol{s}} + \dfrac{4C}{\boldsymbol{s}^2} \\[3ex] \dfrac{dG^-}{dr}\bigg|_{(s/2)} = \dfrac{dG^+}{dr}\bigg|_{(s/2)} & : & \dfrac{\boldsymbol{\rho r s}^2}{\left(1 - \dfrac{1}{6}\boldsymbol{\rho s}^3 \boldsymbol{r}\right)} = -\dfrac{4B}{\boldsymbol{s}^2} - \dfrac{16C}{\boldsymbol{s}^3} \\[3ex] p\boldsymbol{b} = \boldsymbol{r} G(r)\big|_{r \to \infty} & : & \left(1 + \dfrac{2}{3}\boldsymbol{\rho r s}^3 \left(A + B/\boldsymbol{s} + C/\boldsymbol{s}^2\right)\right) = A \end{array}\right\} \tag{87}$$

.

The equations (87) are simply a set of three linear equations in the unknown parameters $A$, $B$, and $C$. The forthright algebraic solutions for $A$, $B$, and $C$, when inserted into the expression for the correlation function, $G(r)$, then give for the pressure equation:

$$p\mathbf{b} = \mathbf{r}\left(\frac{1+\mathbf{h}+\mathbf{h}^2}{(1-\mathbf{h})^3}\right), \quad \text{in which } \mathbf{h} = \frac{p}{6}\mathbf{r}\mathbf{s}^3. \tag{88}$$

Equation (88) is the scaled particle theory equation of state for a hard-sphere fluid.

The foregoing straightforward development of equation (10) via the preceding ones above is clearly exact. The development of Equation (10) by scaled particle theory involves none of the approximations of the Born-Green-Yvon procedure (BGY),[47, 48] and is uncontaminated by the uncertainties of the Hypernetted-Chain-Approximation (HNC)[49] or the Kirkwood approximations.[50, 51] Thus, scaled particle theory provides one of the (*very*) few exact solutions in statistical mechanics.

## Appendix C. Formal development of the description of a chiral hard body, and the separation of the chiral and achiral components of the geometric parameters.

The molecular geometric property of chirality often powerfully effects its thermodynamic behavior. For example, the thermodynamic property that chiral molecules typically condense out of a liquid solution as mixed crystals, *not* as solid solutions, was first demonstrated by Pasteur.

Chirality is a tensor property. The geometric parameters, *R*, *S*, and *V*, which describe a hard-body and enter the Boublík equation of state appear (ostensibly) as scalar entities. The chiral properties of a hard-body particle are described by projecting out of the molecule two different hard-bodies, designated, respectively, the chiral and achiral sub-molecules, as defined below. Each of the geometric parameters, *R*, *S*, and *V*, is then described by a 1×2 matrix:

$$\mathbf{R} = \begin{pmatrix} R^{\text{achiral}} \\ R^{\text{chiral}} \end{pmatrix}, \tag{89}$$

and similarly for *S*.

The factors of the geometric parameters which appear in the Boublík equation of state for mixtures of hard-body molecules, **(16)** or **(18)**, involve bilinear products of *R* and *S*, such as the coefficient of asphericity, $\mathbf{a} = RS/3V$. The bilinear products are defined as the tensorial contraction between the chirality operator, $\mathbf{s}^{\text{chiral}}$, defined in terms of the two-dimensional identity and first Pauli matrices:

$$\begin{pmatrix} 1 & 0 \\ 0 & 1 \end{pmatrix} = \mathbf{I}_2, \quad \begin{pmatrix} 0 & 1 \\ 1 & 0 \end{pmatrix} = \mathbf{s}_1^{\text{Pauli}} \Bigg\}, \tag{90}$$

Because the vector space defining the degree of geometric chirality spans two dimensions, only one of the Pauli matrices are required; and the tensor properties of the geometric functionals are described using the symmetric and anti-symmetric operators:

$$\begin{aligned} \mathbf{s}^S &= \mathbf{I}_2 + \mathbf{s}_1^{\text{Pauli}}, \quad \mathbf{s}^A = 2\mathbf{s}_1^{\text{Pauli}} \\ \mathbf{s}^{\text{chiral}} &= \mathbf{s}^S + \mathbf{s}^A \end{aligned} \Bigg\}. \tag{91}$$

The general bilinear product *RS* is then defined as

$$RS = \left(\sum_i x_i \mathbf{R}_i\right)\left(\mathbf{s}^{\text{Chiral}}\right)\left(\sum_j x_j \mathbf{S}_j\right). \tag{92}$$

One may observe simply that, using equation (92) and its analogous bilinear products in the Boublík equation of state, gives the identical thermodynamic results for pure-component systems of either L or D chiral enantiomers. However, these same equations return additional, different, values for scalemic mixtures of enantiomers, and the Boublík equation of state accordingly returns then, for a given pressure and temperature, a different system vol-

page 28

ume, for which the difference between such and the volume for the pure-component enantiomers is the excess volume.

The chiral and achiral components of the surface of exclusion, $S$, and its mean exterior curvature, $R$, are defined formally as follows.

Let ASO(3) denote the full inhomogeneous rotation group of three dimensional Euclidean space, excluding spatial inversion, and ASO+(3) the group ASO(3) enlarged by including the spatial inversion operation. (The letter "A" stands for affine, which refers to spaces which have no preferred zero point.) For a solid figure in 3-space denoted by F, let ~F denote its image under spatial inversion.

Chiral and achiral figures in 3-space are defined as follows: A figure in 3-space, F, is achiral if there exists $g \in$ ASO(3) such that g(~F) = F. That is, the figure derived from inverting F is identical to F after a suitable combination of translations and rotations. A figure, F, in 3-space is chiral if it is not achiral. That is, F is chiral if it has a handedness, so that it is possible to distinguish it from ~F.

The statement "F is identical to ~F" can mean that there exists $g \in$ ASO(3) which maps the point set of ~F onto the point set of F (which is conventional congruence of geometric figures). Alternately, there exists g $\in$ ASO(3) such that the mapping $g$ ~ (inversion followed by g) is the identify on the point set of F. Therefore $g^{-1}$ and ~ map each part of F to the same part of ~F.

Example: Consider an equilateral tetrahedron, F, which is an achiral figure, in the sense that the point sets of F and ~F are congruent. Now paint the tetrahedron gray, and paint a small region at each vertex red, yellow, green, and blue. F is now chiral, in the sense that F and ~F cannot be placed in space so that corresponding vertices have the same color.

The distinction between equivalence of figures defined by mappings of point-sets and equivalence defined by by mappings that preserve corresponding parts is critical in applications to chemical physics. For example, the tetrahedral molecule methane, $CH_4$, is achiral by any definition, whereas the tetrahedral molecule fluorochloroiodomethane, CHFClI) is achiral when equivalence of figures is defined by mappings that preserve corresponding parts, that is, by mappings that send H to H, F to F, Cl to Cl and I to I. (CHFClI is also chiral when equivalence of figures is defined by point set congruence, but only because the ions have different sizes.)

Henceforth in this paper, equivalence of molecules is defined by mappings that send each radical to another radical of the same type. This situation can be represented by giving identical atoms and ions the same color, and different atoms and ions different colors, and defining equivalence by mappings that preserve colors of atoms and ions.

Definition of an achiral sub-molecule: Given a molecule M, a subset of the radicals in the molecule is an achiral sub-molecule if it is an achiral figure, where equivalence is defined so that radicals of the same type are preserved under equivalence.

Definition: Maximal achiral sub-molecule: Given a molecule M, an achiral sub-molecule is maximal if no other radical of M can be appended to it to yield a larger achiral sub-molecule. An alternate definition could be: The achiral core of a molecule M is the intersection of the maximal achiral sub-molecules of M. The chiral remainder of a molecule M is defined as the complement of the achiral core of M.

Example: For methane, the maximal achiral core is the entire molecule. For CHFClI, the maximal achiral sub-molecules are the carbon plus any two of the "limbs" of the molecule. These sub-molecules are maximal because appending any three of the "limbs" to the carbon results in a chiral molecule. The maximal achiral core of the molecule is the carbon radical. The chiral remainder is the set of radicals {F, Cl, I}, which is not a contiguous molecular segment.

The chiral and achiral components of the geometric parameters $R$, $S$ and $V$, of the achiral core and the chiral remainder of a molecule, which enter the components of the 1×2 matrices indicated by equation (89) are, respectively, defined using differential geometry as follows: The surface parameters, $S$, of the achiral core and of the chiral remainder are both defined as the surface integrals of the two respective parts, but excluding the new surfaces that result from the separation of the two parts.

The mean exterior curvature $R$ of the achiral core and of the chiral remainder are defined, respectively, as the surface integrals of the mean extrinsic curvature over the surfaces of the two parts, but excluding the new surfaces that result from the separation of the two parts. If the seams between the achiral core and the chiral remainder contain any discrete exterior angles (where the surface has a discrete bend), then these discrete exterior angles contribute to the $R$ of the total figure, and a rule would be required for assigning part of this contribution to the achiral core and part to the chiral remainder. However, the surfaces of real molecules do not have any discrete bends.

The parameter $V$ of both the achiral core and of the chiral remainder is trivially defined as the volume of each point set.

These chiral and achiral components of the geometric parameters $R$, $S$ and $V$, were developed for convex hard-body molecules using the cumbersome procedure of Kihara for the analysis of the phenomenon of optical activity in abiotic fluids.[12] In this present case, for which differential geometry has been used, there is no restriction to convex hard-body molecules.